\documentclass[10pt,twocolumn,journal]{IEEEtran}
\usepackage{times}
\usepackage{amsbsy}
\usepackage{latexsym}
\usepackage{amssymb}
\usepackage{amsmath}
\usepackage[usenames]{color}
\usepackage[ansinew]{inputenc}
\usepackage[dvips ]{graphicx}
\input epsf
\usepackage{epic,eepic}

\title{Adaptive Reduced-Rank Equalization Algorithms Based on
Alternating Optimization Design Techniques for Multi-Antenna Systems
}
\author{Rodrigo C. de Lamare and Raimundo Sampaio-Neto
\thanks{This work is partially funded by the
Ministry of Defence (MoD), UK, Contract No. RT/COM/S/021. Part of
this manuscript was presented at WCNC 2008. Dr. R. C. de Lamare is
with the Communications Research Group, Department of Electronics,
University of York, York Y010 5DD, United Kingdom and Prof. R.
Sampaio-Neto is with CETUC/PUC-RIO, 22453-900, Rio de Janeiro,
Brazil. E-mails: rcdl500@ohm.york.ac.uk and
raimundo@cetuc.puc-rio.br}}

\begin{document}
\maketitle

\begin{abstract}
This paper presents a novel adaptive reduced-rank { multi-input
multi-output} (MIMO) equalization scheme and algorithms based on
alternating optimization design techniques for MIMO spatial
multiplexing systems. The proposed reduced-rank equalization
structure consists of a joint iterative optimization of two
equalization stages, namely, a transformation matrix that performs
dimensionality reduction and a reduced-rank estimator that retrieves
the desired transmitted symbol. The proposed reduced-rank
architecture is incorporated into an equalization structure that
allows both decision feedback and linear schemes for mitigating the
inter-antenna and inter-symbol interference. We develop alternating
least squares (LS) expressions for the design of the transformation
matrix and the reduced-rank estimator along with computationally
efficient alternating recursive least squares (RLS) adaptive
estimation algorithms. We then present an algorithm for
automatically adjusting the model order of the proposed scheme. An
analysis of the LS algorithms is carried out along with sufficient
conditions for convergence and a proof of convergence of the
proposed algorithms to the reduced-rank Wiener filter. Simulations
show that the proposed equalization algorithms outperform the
existing reduced-rank and full-rank algorithms, while requiring a
comparable computational cost. \vspace{0.5em}

\end{abstract}

\section{Introduction}

\PARstart{T}{he} high demand for performance and capacity in
wireless networks has led to the development of numerous signal
processing and communications techniques for employing the resources
efficiently. Recent results on information theory have shown that it
is possible to achieve high spectral efficiency~\cite{foschini} and
to make wireless links more reliable \cite{alamouti,tarokh1} through
the deployment of multiple antennas at both transmitter and
receiver. In { multi-input multi-output} (MIMO) communications
systems, the received signal is composed by the sum of several
transmitted signals which share the propagation environment and are
subject to { multi-path propagation effects} and noise at the
receiver. The multipath channel originates inter-symbol interference
(ISI), whereas the non-orthogonality among the signals transmitted
gives rise to inter-antennas interference (IAI) at the receiver.

In order to mitigate the effects of ISI and IAI that reduce the
performance and the capacity of MIMO systems the designer has to
construct a MIMO equalizer. The optimal MIMO equalizer known as the
maximum likelihood sequence estimation (MLSE) receiver was
originally developed in the context of multiuser detection in
\cite{verdu}. However, the exponential complexity of the optimal
MIMO equalizer makes its implementation costly for multipath
channels with severe ISI and MIMO systems with many antennas. In
practice, designers often prefer the deployment of low-complexity
MIMO receivers such as the linear \cite{duel,delamareccmmimo}, the
successive interference cancellation-based VBLAST \cite{vblast} and
decision feedback equalizers (DFE) \cite{ginis}-\cite{choi}. The DFE
schemes \cite{ginis}-\cite{choi} can achieve significantly better
performance than linear ones due to the interference cancellation
capabilities of the feedback section. These receivers require the
estimation of the coefficients used for combining the received data
and extracting the desired transmitted symbols. A challenging
problem in MIMO systems \cite{hassibi} { is encountered} when the
number of elements in the equalizer or the number of antenna pairs
is large, which is key to future applications
\cite{chiu}-\cite{mohammed}. In these situations, an estimation
algorithm requires substantial training for the MIMO equalizer and a
large number of received symbols to reach its steady-state behavior.

There are many algorithms for designing MIMO equalizers, which
possess different trade-offs between performance and complexity
\cite{haykin}. In this regard, least squares (LS)-based algorithms
are often the preferred choice with respect to convergence
performance. However, when the number of filter elements in the
equalizer is large, { an adaptive LS-type algorithm} requires a
large number of samples to reach its steady-state behavior and may
encounter problems in tracking the desired signal. Reduced-rank
techniques \cite{scharfo}-\cite{jidf} are powerful and effective
approaches in low-sample support situations and in problems with
large filters. These algorithms can exploit the low-rank nature of
signals that are found in MIMO communications \cite{gesbert} in
order to achieve faster convergence speed, increased robustness
against interference and better tracking performance than full-rank
techniques. By projecting the input data onto a low-rank subspace
associated with the signals of interest, reduced-rank methods can
eliminate the interference that lies in the noise subspace and
perform denoising \cite{scharfo}-\cite{jidf}. Prior work on
reduced-rank estimators for MIMO systems is extremely limited and
relatively unexplored, being the work of Sun \textit{et al.}
\cite{sun} one of the few existing ones in the area. A comprehensive
study of reduced-rank equalization algorithms for MIMO systems has
not been considered so far.  It is well known that the optimal
reduced-rank approach is based on the eigen-decomposition (EVD) of
the known input data covariance matrix ${\boldsymbol R}$
\cite{scharfo}. However, this covariance matrix must be estimated.
The approach taken to estimate ${\boldsymbol R}$ and perform
dimensionality reduction is of central importance and plays a key
role in the performance of the system. Numerous reduced-rank
strategies have been proposed in the last two decades. The first
methods were based on the EVD of time-averaged estimates of
${\boldsymbol R}$ \cite{scharfo}, in which the dimensionality
reduction is carried out by a transformation matrix formed by
appropriately selected eigenvectors computed with the EVD. A more
recent and elegant approach to the problem was taken with the advent
of the multistage Wiener filter (MSWF) \cite{gold&reed}, which was
later extended to adaptive versions in \cite{goldstein,mswfccm}, and
MIMO applications \cite{sun}. Another related method is the
auxiliary vector filtering (AVF) algorithm \cite{avf3}-\cite{avf5},
which can outperform the MSWF. A key limitation with prior art is
the deficient exchange of information between the dimensionality
reduction task and the subsequent reduced-rank estimation.

In this work, we propose adaptive reduced-rank MIMO equalization
algorithms based on alternating optimization design techniques for
MIMO spatial multiplexing systems. The proposed reduced-rank
equalization structure and algorithms consist of a joint iterative
optimization that alternates between two equalization stages,
namely, a transformation matrix that performs dimensionality
reduction and a reduced-rank estimator that suppresses the IAI
caused by the associated data streams and retrieves the desired
transmitted symbol. The essence of the proposed scheme is to
change the role of the equalization filters and promote the
exchange of information between the dimensionality reduction and
the reduced-rank estimation tasks in an alternated way. In order
to estimate the coefficients of the proposed MIMO reduced-rank
equalizers, we develop alternating least squares (LS) optimization
algorithms and expressions for the joint design of the
transformation matrix and the reduced-rank filter. We derive
alternating recursive LS (RLS) adaptive algorithms for their
computationally efficient implementation and present a complexity
study of the proposed and existing algorithms. We also describe an
algorithm for automatically adjusting the model order of the
proposed reduced-rank MIMO equalization schemes. An analysis of
the proposed LS optimization is conducted, in which sufficient
conditions and proofs for the convergence of the proposed
algorithms are derived. The performance of the proposed scheme is
assessed via simulations for MIMO equalization applications.
The main contributions of this work are summarized as follows: \\
1) A reduced-rank MIMO equalization scheme and a design approach
for both decision feedback and linear structures;
\\ 2) Reduced-rank LS expressions and recursive algorithms for parameter
estimation; \\ 3) An algorithm for automatically adjusting the
model order;\\  4) Analysis and convergence proofs of the proposed
algorithms.  \\ 5) A study of MIMO reduced-rank equalization
algorithms.

This paper is structured as follows: The MIMO system { and signal
model is described} in Section II. The proposed adaptive MIMO
reduced-rank equalization structure is introduced along with the
problem statement in Section III. Section IV is devoted to the
development of the LS estimators, the computationally efficient RLS
algorithms and the model order selection algorithms. Section V
presents an analysis and proofs of convergence of the proposed
algorithms. Section VI discusses the simulation results and Section
VII gives the conclusions of this work.

{ \emph{Notation:} In this paper bold upper and lowercase letters
represent matrices and vectors, respectively. $(.)^*$, $(.)^*H$,
$(.)^{-1}$ and $(.)^T$ shall represent complex conjugate, complex
conjugate transpose (Hermitian), inverse and transpose,
respectively. ${\rm tr}(.)$ is the trace operator of a matrix.
Reduced-rank vectors and matrices are given with the addition of a
bar $(\bar{.})$ and estimated symbols are denoted by the addition of
a hat $(\hat{.})$.}

\section{MIMO System and Signal Model}

In this section we present { MIMO communications system and signal
model and describe its} main components. The model in this section
is intended for describing a general MIMO system in multipath
channels. However, it can also serve as a model for broadband MIMO
communications systems with guard intervals including those based on
orthogonal frequency-division multiplexing (OFDM) \cite{li,stuber}
and single-carried (SC) modulation with frequency-domain
equalization \cite{falconer}. 

\begin{figure}[!htb]
\begin{center}
\def\epsfsize#1#2{1.0\columnwidth}
\epsfbox{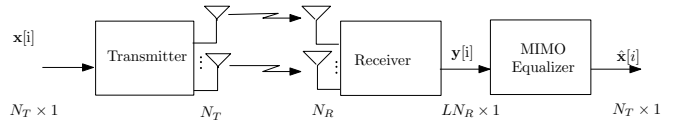} \caption{ MIMO system model.}\label{fig:system}
\end{center}
\end{figure}

Consider a MIMO system with $N_T$ antennas at the transmitter and
$N_R$ antennas at the receiver in a spatial multiplexing
configuration, as shown in Fig.~\ref{fig:system}. {  The system is
mathematically equivalent to that in \cite{dhahir}.} The signals are
modulated and transmitted from $N_T$ antennas over multipath
channels whose propagation effects are modelled by finite impulse
response (FIR) filters with $L_p$ coefficients, and are received by
$N_R$ antennas. We assume that the channel can vary during each
packet transmission and the receiver is perfectly synchronized with
the main propagation path. At the receiver, a MIMO equalizer is used
to mitigate IAI and ISI and retrieve the transmitted signals.

The signals transmitted by the system at time instant $i$ can be
described by ${\boldsymbol x}[i] = [x_1[i] ~\ldots~
x_{N_T}[i]]^T$, where $x_j[i]$, $j=1, ~\ldots~N_T$ are independent
and identically distributed symbols of unit variance. The
demodulated signal received at the $k$th antenna and time instant
$i$ after applying a filter matched to the signal waveform and
sampling at symbol rate is expressed by
\begin{equation}
y_k[i] = \sum_{j=1}^{N_T} \sum_{l=0}^{L_p-1} h_{j,k,l}[i] x_j[i-l]
+ n_k[i], ~~~{\rm for}~~ k = 1, \ldots, N_R,
\end{equation}
where $h_{j,k,l}[i]$ is the sampled impulse response between
transmit antenna $j$ and receive antenna $k$ for path $l$, and
$n_k[i]$ are samples of white Gaussian complex noise with zero mean and variance
$\sigma^2$. By collecting the samples of the received signal and
organizing them in a window of $L$ symbols ($L \geq L_p$) for each
antenna element, we obtain the $L N_R \times 1$ received vector
\begin{equation}
\label{eq:one} {\boldsymbol y}[i] = {\boldsymbol H}[i]
{\boldsymbol x}_T[i]  + {\boldsymbol n}[i],
\end{equation}
where ${\boldsymbol y}[i] = \big[ {\boldsymbol y}_1^T[i] ~ \ldots
~{\boldsymbol y}_{N_R}^T[i] \big]^T$ contains the signals
collected by the $N_R$ antennas, the $L \times 1$ vector
${\boldsymbol y}_k[i] = \big[ y_{k}[i] ~ \ldots ~ y_{k}[i-L+1]
\big]^T$, for $k=1, ~\ldots,~ N_R$, contains the signals collected
by the $k$th antenna and are organized into a vector. The window
size $L$ must be chosen according to the prior knowledge about the
delay spread of the multipath channel \cite{rappa}. The $L N_R
\times L N_T$ MIMO channel matrix ${\boldsymbol H}[i]$ is
\begin{equation}
\label{eq:two} {\boldsymbol H}[i] = \left[ \begin{array}{cccc}
  {\boldsymbol H}_{1,1}[i] & {\boldsymbol H}_{1,2}[i] & \ldots & {\boldsymbol H}_{1,N_T}[i] \\
  {\boldsymbol H}_{2,1}[i] & {\boldsymbol H}_{2,2}[i] & \ldots & {\boldsymbol H}_{2,N_T}[i] \\
  \vdots & \vdots & \ddots & \vdots \\
  {\boldsymbol H}_{N_R,1}[i] & {\boldsymbol H}_{N_R,2}[i] & \ldots & {\boldsymbol H}_{N_R,N_T}[i]
\end{array}\right],
\end{equation}
{  where the $L \times L$ matrix ${\boldsymbol H}_{j,k}[i]$ are
Toeplitz matrices with the channel gains organized in a channel
vector ${\boldsymbol h}_{j,k}[i] = [h_{j,k,1}[i] ~ \ldots ~
h_{j,k,L_p-1}]^T$ that is shifted down by one position from left to
right for each column, and which describes the multi-path channel
from antenna $j$ to antenna $k$. The elements $h_{j,k,l}[i]$, for
$l=0, ~ \ldots,~ L_p$, of ${\boldsymbol h}_{j,k}[i]$ are modelled as
random variables and follow a specific propagation channel model
\cite{rappa}, as will be detailed in the Section VI.} The $L N_T
\times 1$ vector ${\boldsymbol x}_T[i] = \big[ {\boldsymbol
x}_1^T[i] ~ \ldots ~ {\boldsymbol x}_{N_T}^T[i] \big]^T$ is composed
by the data symbols transmitted from the $N_T$ antennas at the
transmitter with ${\boldsymbol x}_{j}[i] = [x_j[i] ~\ldots
~x_j[i-L+1]]^T$ being the $i$th transmitted block with dimensions $L
\times 1$. The $L N_R \times 1$ vector ${\boldsymbol n}[i]$ is a
complex Gaussian noise vector with zero mean and $E \big[
{\boldsymbol n}[i] {\boldsymbol n}^H[i] \big] = \sigma^2
{\boldsymbol I}$, where $(\cdot)^{T}$ and $(\cdot)^{H}$ denote
transpose and Hermitian transpose, respectively, and $E[\cdot]$
stands for expected value.

\section{Proposed Adaptive Reduced-Rank MIMO DFE and Problem Formulation}

\begin{figure*}[t]
\begin{center}
\def\epsfsize#1#2{1.8\columnwidth}
\epsfbox{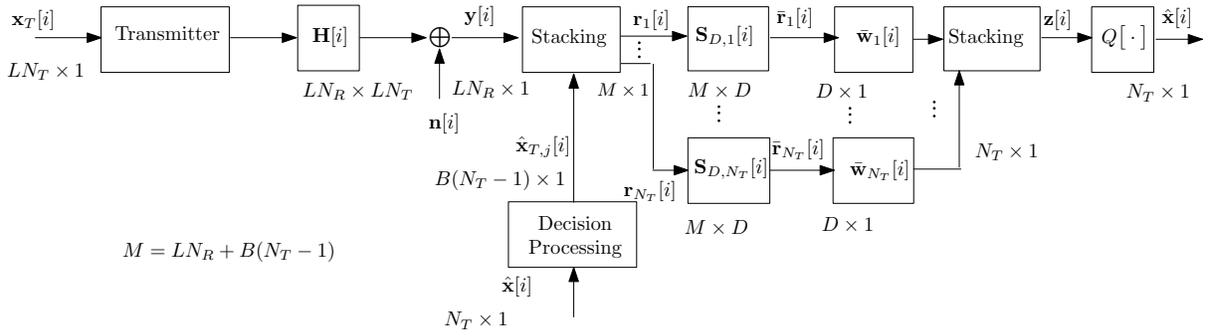} \caption{Proposed MIMO reduced-rank decision
feedback equalization structure.} \label{pscheme}
\end{center}
\end{figure*}

We present the proposed reduced-rank MIMO equalization structure
and state the main design problem of reduced-rank MIMO
equalization structures. Both decision feedback (DF) and linear
equalization structures can be devised by adjusting the dimensions
of the filters and the use of feedback. We shall start with the
description of the DF structure and then obtain the linear scheme
as a particular case. In the proposed MIMO reduced-rank DF
equalizer (DFE), the signal processing tasks are carried out in
two stages, as illustrated in Fig. \ref{pscheme}. The proposed
scheme employs two sets of filters and stacks the decision and the
input data vectors for joint processing. The decision feedback
strategy adopted in this work is the parallel scheme reported in
\cite{dhahir,delamaretc}, which firstly obtains the decision
vector $\hat{{\boldsymbol x}}_{T,j}[i]$ with linear equalization
and then employs $\hat{{\boldsymbol x}}_{T,j}[i]$ to cancel the
interference caused by the interfering streams. A decision delay
$\delta_{\rm dec}$ is assumed between the symbols transmitted and
the $\hat{{\boldsymbol x}}_{T,j}[i]$ obtained after the decision
block. The parallel strategy outperforms the successive one that
uses a sequential procedure of equalization and interference
cancellation \cite{vblast,ginis}.

Let us consider the design of the proposed MIMO reduced-rank
equalizer using the structure shown in Fig. \ref{pscheme}. The $M
\times 1$ input data vector ${\bf r}[i]$ to the proposed equalizer
is obtained by stacking the $LN_R \times 1$ received vector
${\boldsymbol y}[i]$ and the $B(N_T-1) \times 1$ vector of
decisions $\hat{\boldsymbol x}_{T,j}[i]$ for stream $j$ and is
described by
\begin{equation}
{\boldsymbol r}_j[i] = \left[
\begin{array}{c} {\boldsymbol y}[i] \\ \hat{{\boldsymbol x}}_{T,j}[i]
\end{array} \right], \label{rj}
\end{equation}
where $M=LN_R+B(N_T-1)$ represents the number of samples for
processing. The $B(N_T-1) \times 1$ vector of decisions
$\hat{{\boldsymbol x}}_{T,j}[i]= [\hat{\boldsymbol x}_j[i]~ \ldots
~\hat{\boldsymbol x}_j[i-B+1]]^T$ for the $j$th stream takes into
account $B$ decision instants for the feedback and excludes the
$j$th detected symbol to avoid cancelling the desired symbol. The
$N_T \times 1$ vector of decisions is given by $\hat{\boldsymbol
x}[i] = [\hat{x}_1[i]~ \ldots ~\hat{x}_{N_T}[i]]^T $, whereas the
$N_T-1 \times 1$ vector of decisions that excludes stream $j$ and
is employed to build $\hat{{\boldsymbol x}}_{T,j}[i]$ is given by
$\hat{\boldsymbol x}_j[i] = [\hat{x}_1[i]~ \ldots ~
\hat{x}_{j-1}[i] ~\hat{x}_{j+1}[i]~ \ldots ~\hat{x}_{N_T}[i]]^T $.

Let us now consider an $M \times D$ transformation matrix
${\boldsymbol S}_{D,j}[i]$ which carries out a dimensionality
reduction on the received data ${\boldsymbol r}_j[i]$ and { shall}
exploit the low-rank nature of the data transmitted over stream $j$
as follows
\begin{equation}
{\bar{\boldsymbol r}}_j[i] = {\boldsymbol S}_{D,j}^H[i]
{\boldsymbol r}_j[i], ~~ j=1, \ldots, N_T,
\end{equation}
where $D$ is the rank of the resulting equalization system.

The resulting projected received vector ${\bar{\boldsymbol
r}}_j[i]$ is the input to an estimator represented by the $D
\times 1$ vector ${\bar{\boldsymbol w}}_j[i]=[ {\bar
w}_{j,1}^{}[i] ~{\bar w}_{j,2}^{}[i]~\ldots ~ {\bar
w}_{j,D}^{}[i]]^T$. According to the schematic shown in Fig.
\ref{pscheme}, the output of the proposed MIMO reduced-rank DFE is
obtained by linearly combining the coefficients of ${\boldsymbol
S}_{D,j}[i]$ and ${\bar{\boldsymbol w}}_j[i]$ for extracting the
symbol transmitted from antenna $j$. Notice that all
$D$-dimensional quantities have a "bar". The proposed MIMO
reduced-rank DFE output is
\begin{equation}
\begin{split}
\tilde{z}_j[i] & = {\bar{\boldsymbol w}}^{H}_j[i] {\boldsymbol
S}_{D,j}^H[i] {\boldsymbol r}_j[i]  = {\bar{\boldsymbol
w}}^{H}_j[i]{\bar{\boldsymbol r}}_j[i],
\end{split}
\end{equation}
From the outputs $z_j[i]$ for $j=1,~2, \ldots, N_T$, we construct
the vector ${\boldsymbol z}[i] = \big[ z_1[i] \ldots z_j[i] \ldots
z_{N_T}[i] \big]^T$. The initial decisions for each data stream
are obtained without resorting to the feedback and are computed as
follows
\begin{equation}\hat{ x}_j[i] = Q \Big( \bar{\boldsymbol w}^{H}_j[i] {\boldsymbol
S}_{D,j}^H[i] \left[
\begin{array}{c} {\boldsymbol y}[i] \\ {\boldsymbol 0}
\end{array} \right]  \Big),
\end{equation}
where $Q \big( \cdot \big)$ represents a decision device suitable
for the constellation of interest (BPSK, QPSK or QAM) and the vector
of decisions is constructed as $\hat{\boldsymbol x}[i] = \big[ {\hat
x}_1[i] \ldots {\hat x}_j[i] \ldots {\hat x}_{N_T}[i] \big]^T$ and
used to construct $\hat{{\boldsymbol x}}_{T,j}[i]$ and ${\boldsymbol
r}_j[i]$ as in (\ref{rj}). The detected symbols $\hat{\boldsymbol
x}^{(f)}[i]$ of the proposed reduced-rank MIMO DFE after the IAI and
ISI cancellation are obtained by
\begin{equation}
\hat{\boldsymbol x}^{(f)}[i] = Q \big( {\boldsymbol z}[i] \big)= Q
\Bigg( \left[\begin{array}{c} {\bar{\boldsymbol w}}^{H}_1[i]
{\boldsymbol S}_{D,1}^H[i] {\boldsymbol r}_1[i] \\ \vdots \\
{\bar{\boldsymbol w}}^{H}_{N_T}[i] {\boldsymbol S}_{D,N_{T}}^H[i]
{\boldsymbol r}_{N_T}[i]\end{array} \right] \Bigg).
\end{equation}
The feedback employs $B(N_T-1)$ connections for cancelling the IAI
and the other $N_T-1$ data streams and the ISI from the adjacent
symbols. A reduced-rank MIMO linear equalizer is obtained by
neglecting the feedback with decision processing of the structure
in Fig. \ref{pscheme}.

The previous development suggests that the key aspect and problem
to be solved in the design of reduced-rank MIMO equalization
schemes is the cost-effective computation of the estimators
${\boldsymbol S}_{D,j}[i]$ and $\bar{\boldsymbol w}_j[i]$. The
transformation matrix ${\boldsymbol S}_{D,j}[i]$ plays the most
important role since it carries out the dimensionality reduction,
which profoundly affects the performance of the remaining
estimators and the MIMO equalizers. Methods based on the EVD
\cite{scharfo}, the MSWF
\cite{goldstein} and the AVF \cite{avf3}-\cite{avf5} were
reported for the design of ${\boldsymbol S}_{D,j}[i]$, however,
they did not consider jointly the design of ${\boldsymbol
S}_{D,j}[i]$ and $\bar{\boldsymbol w}_j[i]$ via alternating
optimization recursions. In the next section, we present the
reduced-rank least squares (LS) algorithms and their recursive
versions for the design of ${\boldsymbol S}_{D,j}[i]$ and
$\bar{\boldsymbol w}_j[i]$ used in the proposed MIMO equalization
structure.

\section{Proposed Reduced-Rank Least Squares Design and Adaptive Algorithms}

In this section, we present a joint iterative exponentially weighted
reduced-rank LS estimator design of the parameters ${\boldsymbol
S}_{D,j}[i]$ and $\bar{\boldsymbol w}_j[i]$ of the proposed MIMO
reduced-rank DFE. We then derive computationally efficient
algorithms for computing the proposed LS estimator in a recursive
way and automatically adjusting the model order. The deficient
exchange of information between the dimensionality reduction task
and the reduced-rank estimation verified in previously reported
algorithms \cite{gold&reed}-\cite{avf5} is addressed by the
alternated procedure that updates ${\boldsymbol S}_{D,j}[i]$ and
$\bar{\boldsymbol w}_j[i]$. {  Specifically, the expression of
${\boldsymbol S}_{D,j}[i]$ is a function of $\bar{\boldsymbol
w}_j[i]$ and vice versa, and this allows the coefficients to be
computed via an alternating procedure with exchange of information
in both ways (from ${\boldsymbol S}_{D,j}[i]$ to $\bar{\boldsymbol
w}_j[i]$ and the other way around). Our studies and numerical
results indicate that this approach is more effective than the MSWF
\cite{goldstein} and the AVF  \cite{avf5} algorithms. In addition,
the rank reduction is based on the joint and iterative LS
minimization which has been found superior to the Krylov subspace,
as evidenced in the numerical results. This allows the proposed
method to outperform the MSWF and the AVF. }We have opted for the
use of one cycle (or iteration) per time instant in order to keep
the complexity low. We also detail the computational complexity of
the proposed and existing algorithms in terms of arithmetic
operations.

\subsection{Reduced-Rank Least Squares Estimator Design}

In order to design ${\boldsymbol S}_{D,j}[i]$ and
$\bar{\boldsymbol w}_j[i]$, we describe a joint iterative
reduced-rank LS optimization algorithm. Consider the
exponentially-weighted LS expressions for the estimators
${\boldsymbol S}_{D,j}[i]$ and $\bar{\boldsymbol w}_j[i]$ via the
cost function
\begin{equation}
\begin{split}
{\mathcal{C}}_j({\boldsymbol S}_{D,j}[i],~\bar{\boldsymbol w}_j[i]
) & = \sum_{l=1}^i \lambda^{i-l} \big| x_j[l] - \bar{\boldsymbol
w}^H_j[i] {\boldsymbol S}_{D,j}^H[i] {\boldsymbol r}_j[l] \big|^2,
\label{costfunc}
\end{split}
\end{equation}
where $0<\lambda \leq 1$ is the forgetting factor.

The proposed exponentially-weighted LS design corresponds to
solving the following optimization problem
\begin{equation}
\begin{split}
\big\{ {\boldsymbol S}_{D,j}^{\rm opt}, ~ \bar{\boldsymbol
w}_{j}^{\rm opt} \big\} = \arg \min_{\bar{{\boldsymbol
S}_{D,j}[i],\boldsymbol w}_j[i]} {\mathcal{C}}_j ({\boldsymbol
S}_{D,j}[i],~\bar{\boldsymbol w}_j[i]) \label{opt_prob}
\end{split}
\end{equation}
In order to solve the problem in (\ref{opt_prob}), the proposed
strategy is to fix a set of parameters, find the other set of
parameters that minimize (\ref{costfunc}) and alternate this
procedure between the two sets ${\boldsymbol S}_{D,j}[i]$ and
$\bar{\boldsymbol w}_j[i]$. By minimizing (\ref{costfunc}) with
respect to ${\boldsymbol S}_{D,j}[i]$, we obtain
\begin{equation}
{\boldsymbol S}_{D,j}[i] = {\boldsymbol R}^{-1}_j[i] {\boldsymbol
P}_{D,j}[i]  {\boldsymbol R}_{\bar{{\boldsymbol
w}}_j}^{\dagger}[i-1],\label{filtersd}
\end{equation}
where the $M \times D$ matrix ${\boldsymbol P}_{D,j}[i] =
\sum_{l=1}^i \lambda^{i-l}x^{*}_j[l]{\boldsymbol
r}_j[l]\bar{\boldsymbol w}^H_j[i-1]$, ${\boldsymbol R}_j[i] =
\sum_{l=1}^i \lambda^{i-l}{\boldsymbol r}_j[l]{\boldsymbol
r}_j^{H}[l]$, {  $(\cdot)^{\dagger}$ denotes the Moore-Penrose
pseudo-inverse} and the $D \times D$ matrix ${\boldsymbol R}_{{\bar
w}_j}[i-1] = \bar{\boldsymbol w}_j[i-1]\bar{\boldsymbol
w}^{H}_j[i-1]$. Since ${\boldsymbol R}_{{\bar w}_j}[i-1]$ is a
rank-$1$ matrix, we need to either compute the pseudo-inverse or
introduce a regularization term in the recursion ${\boldsymbol
R}_{{\bar w}_j}[i-1] = \sum_{l=1}^{i-1}
\lambda^{i-l}{\bar{\boldsymbol w}}_j[l]{\bar{\boldsymbol
w}}^{H}_j[l]$. We have opted for using the latter with the initial
regularization factor ${\boldsymbol R}_{{\bar w}_j}[0]=\delta
{\boldsymbol I}$ for numerical and { simplicity} reasons.

By minimizing (\ref{costfunc}) with respect to $\bar{\boldsymbol
w}_j[i]$, the reduced-rank estimator becomes
\begin{equation}
\bar{\boldsymbol w}_j[i] = \bar{\boldsymbol R}^{-1}_j[i]
\bar{\boldsymbol p}_j[i], \label{filterw}
\end{equation}
where $\bar{\boldsymbol p}_j[i] = {\boldsymbol S}_{D,j}^H[i]
\sum_{l=1}^i \lambda^{i-l}x^{*}_j[l]{\boldsymbol r}_j[l] =
\sum_{l=1}^i \lambda^{i-l}x^{*}_j[l]\bar{\boldsymbol r}_j[l]]$,
and the $D \times D$ reduced-rank correlation matrix is described
by $\bar{\boldsymbol R}_j[i] = {\boldsymbol
S}_{D,j}^H[i]\sum_{l=1}^i \lambda^{i-l}{\boldsymbol
r}_j[l]{\boldsymbol r}_j^H[l] {\boldsymbol S}_{D,j}[i] $.

The equation with the associated sum of error squares (SES) { is
obtained by substituting the expressions in (\ref{filtersd}) and
(\ref{filterw}) into the cost function (\ref{costfunc}),} and is
given by
\begin{equation}
\begin{split}
{\rm SES} & = \sigma^{2}_{x_j}  - \bar{\boldsymbol w}_j^H[i]
{\boldsymbol S}_{D,j}^H[i] {\boldsymbol p}[i] - {\boldsymbol
p}^H[i]{\boldsymbol S}_{D,j}[i]\bar{\boldsymbol w}_j[i] \\ & \quad
+ \bar{\boldsymbol w}^H[i] {\boldsymbol S}_{D,j}^H[i] {\boldsymbol
R}_j[i]{\boldsymbol S}_{D,j}[i] \bar{\boldsymbol w}_j[i] ,
\label{ses}
\end{split}
\end{equation}
where $\sigma^{2}_{x_j}=\sum_{l=1}^i \lambda^{i-l}|x_j[l]|^{2}$.
Note that the expressions in (\ref{filtersd}) and (\ref{filterw})
are not closed-form solutions for $\bar{\boldsymbol w}_j[i]$ and
${\boldsymbol S}_{D,j}[i]$ since they depend on each other and,
thus, they have to be alternated with an initial guess to obtain a
solution. The key strategy lies in the joint optimization of the
estimators. The rank $D$ must be set by the designer to ensure
appropriate performance. The computational complexity of
calculating (\ref{filtersd}) and (\ref{filterw}) is cubic with the
number of elements in the estimators, namely, $M$ and $D$,
respectively. In what follows, we introduce efficient RLS
algorithms for computing the estimators with a quadratic cost.

\subsection{Reduced-Rank RLS Algorithms}

In this part, we present a recursive approach for efficiently
computing the LS expressions developed in the previous subsection.
Specifically, we develop reduced-rank RLS algorithms for computing
$\bar{\boldsymbol w}_j[i]$ and ${\boldsymbol S}_{D,j}[i]$. Unlike
conventional (full-rank) RLS algorithms that require the calculation
of one estimator for the MIMO DFE, the proposed reduced-rank RLS
technique jointly and iteratively computes the transformation matrix
${\boldsymbol S}_{D,j}[i]$ and the reduced-rank estimator
$\bar{\boldsymbol w}_j[i]$. In order to start the derivation of the
proposed algorithms, let us define
\begin{equation}
\begin{split}
{\boldsymbol  P}_j[i] & \triangleq {\boldsymbol R}^{-1}_j[i], \\
{\boldsymbol Q}_{\bar{\boldsymbol w}_j}[i-1] & \triangleq
{\boldsymbol
R}^{-1}_{\bar{\boldsymbol w}_j}[i-1], \\
{\boldsymbol P}_{D,j} [i] & \triangleq  \lambda {\boldsymbol
P}_{D,j}[i-1] + x^*_j[i] {\boldsymbol r}_j[i] \bar{\boldsymbol
w}^H [i-1]. \label{defmat}
\end{split}
\end{equation}
Rewriting the expression in (\ref{filtersd}), we arrive at
\begin{equation}
\begin{split}
{\boldsymbol S}_{D,j} [i] & = {\boldsymbol R}^{-1}_j[i]
{\boldsymbol P}_{D,j} [i]  {\boldsymbol R}_{\bar{\boldsymbol w}_j}^{-1}[i-1] \\
& = {\boldsymbol P}_j[i] {\boldsymbol P}_{D,j} [i] {\boldsymbol Q}_{\bar{\boldsymbol w}_j}[i-1] \\
& = {\boldsymbol S}_{D,j}[i-1] + {\boldsymbol k}_j[i] \big( {
x}_j^*[i] {\boldsymbol t}_j^H[i-1] \\ & \quad - {\boldsymbol
r}_j^H[i]{\boldsymbol S}_{D,j}[i-1]  \big),
\end{split}
\end{equation}
where the $D \times 1$ vector ${\boldsymbol t}_j[i-1] =
{\boldsymbol Q}_{\bar{\boldsymbol w}_j}[i-1] \bar{\boldsymbol
w}_j[i-1]$, the $M \times 1$ Kalman gain vector is
\begin{equation}
{\boldsymbol k}_j[i] = \frac{\lambda^{-1} {\boldsymbol P}_j[i-1]
{\boldsymbol r}_j[i] }{1 + \lambda^{-1} {\boldsymbol r}_j^H[i]
{\boldsymbol P}_j[i-1]{\boldsymbol r}_j[i]},
\end{equation}
the update for the $M \times M$ matrix ${\boldsymbol P}_j[i]$
employs the matrix inversion lemma \cite{haykin}
\begin{equation}
{\boldsymbol P}_j[i] = \lambda^{-1} {\boldsymbol P}_j[i-1] -
\lambda^{-1} {\boldsymbol k}_j[i] {\boldsymbol r}_j^H[i]
{\boldsymbol P}_j[i-1],
\end{equation}
and the $D \times 1$ vector ${\boldsymbol t}_j[i-1]$ is updated as
\begin{equation}
{\boldsymbol t}_j[i-1] = \frac{\lambda^{-1} {\boldsymbol
Q}_{\bar{\boldsymbol w}_j}[i-1] \bar{\boldsymbol w}_j[i-1] }{1 +
\lambda^{-1} \bar{\boldsymbol w}^H_j[i-1] {\boldsymbol
Q}_{\bar{\boldsymbol w}_j}[i-1]\bar{\boldsymbol w}_j[i-1]}.
\end{equation}
The matrix inversion lemma is then used to update the $D \times D$
matrix ${\boldsymbol Q}_{\bar{\boldsymbol w}_j}[i-1]$ as described
by
\begin{equation}
{\boldsymbol Q}_{\bar{\boldsymbol w}_j}[i-1] = \lambda^{-1}
{\boldsymbol Q}_{\bar{\boldsymbol w}_j}[i-2] - \lambda^{-1}
{\boldsymbol t}_j[i-1] \bar{\boldsymbol w}^H_j[i-2]{\boldsymbol
Q}_{\bar{\boldsymbol w}_j}[i-2]. \label{Qmat}
\end{equation}
Equations (\ref{defmat})-(\ref{Qmat}) constitute the part of the
proposed reduced-rank RLS algorithms for computing ${\boldsymbol
S}_{D,j}[i]$.

In order to develop the second part of the algorithm that
estimates $\bar{\boldsymbol w}_j[i]$, let us consider the
expression in (\ref{filterw}) with its associated quantities,
i.e., the $D \times D$ matrix $\bar{\boldsymbol R}_j[i] =
\sum_{l=1}^i \lambda^{i-l}\bar{\boldsymbol r}_j[l]
\bar{\boldsymbol r}_j^{H}[l]$ and the $D \times 1$ vector
$\bar{\boldsymbol p}_j[i] = \sum_{l=1}^i
\lambda^{i-l}x^{*}_j[l]\bar{\boldsymbol r}_j[l]$.

Let us now define $\boldsymbol{\bar{\Phi}}_j[i] = {\boldsymbol
R}^{-1}_j[i]$ and rewrite $\bar{\boldsymbol p}_j[i]$ as
$\bar{\boldsymbol p}_j[i] = \lambda \bar{\boldsymbol p}_j[i-1] +
x^*_j[i] \bar{\boldsymbol r}_j[i]$. We can then rewrite
(\ref{filterw}) as follows
\begin{equation}
\begin{split}
\bar{\boldsymbol w}_j[i] & = \boldsymbol{\bar{\Phi}}_j[i]
\bar{\boldsymbol p}_{j}[i]  \\
& = \bar{\boldsymbol w}_j[i-1] - \bar{\boldsymbol k}_j[i]
\bar{\boldsymbol r}^H_j[i]
\bar{\boldsymbol w}_j[i-1]  + \bar{\boldsymbol k}_j[i] x^*_j[i]   \\
& = \bar{\boldsymbol w}_j[i-1] + \bar{\boldsymbol k}_j[i] \big(
x^*_j[i] - \bar{\boldsymbol r}^H_j[i] \bar{\boldsymbol w}_j[i-1]
\big).
\end{split}
\end{equation}
By defining $\xi_j[i] = x_j[i] - \bar{\boldsymbol
w}^H_j[i-1]\bar{\boldsymbol r}_j[i]$ we arrive at the proposed RLS
algorithm for computing $\bar{\boldsymbol w}_j[i]$
\begin{equation}
\bar{\boldsymbol w}_j[i] = \bar{\boldsymbol w}_j[i-1] +
\bar{\boldsymbol k}_j[i] \xi^*_j[i], \label{wpart}
\end{equation}
where the $D \times 1$ Kalman gain vector is given by
\begin{equation}
\bar{\boldsymbol k}_j[i] = \frac{\lambda^{-1}
\boldsymbol{\bar{\Phi}}_j[i-1] \bar{\boldsymbol r}_j[i] }{1 +
\lambda^{-1} \bar{\boldsymbol r}^H_j[i]
\boldsymbol{\bar{\Phi}}_j[i-1]\bar{\boldsymbol r}_j[i]},
\end{equation}
and the update for the matrix inverse $\boldsymbol{\bar{\Phi}}[i]$
employs the matrix inversion lemma \cite{haykin}
\begin{equation}
\boldsymbol{\bar{\Phi}}_j[i] = \lambda^{-1}
\boldsymbol{\bar{\Phi}}_j[i-1] - \lambda^{-1} \bar{\boldsymbol
k}_j[i] \bar{\boldsymbol r}^H_j[i] \boldsymbol{\bar{\Phi}}_j[i-1].
\label{phipart}
\end{equation}
Equations (\ref{wpart})-(\ref{phipart}) constitute the second part
of the proposed algorithm that computes $\bar{\boldsymbol
w}_j[i]$. The computational complexity of the proposed RLS
algorithms is $O(D^2)$ for the estimation of $\bar{\boldsymbol
w}_j[i]$ and $O(M^2)$ for the estimation of ${\boldsymbol
S}_{D,j}[i]$. Since $D << M$ for moderate to large $L$, $N_R$,
$N_T$ and $B$, as will be explained in the next section, the
overall complexity is in the same order of the conventional
full-rank RLS algorithm ($O(M^2)$) \cite{haykin}.

\subsection{Model-Order Selection Algorithm}

The performance of the LS { and} RLS algorithms described in the
previous subsection depends on the model order or the rank $D$. This
motivates the development of methods to automatically adjust $D$
using an LS cost function as a mechanism to control the selection.
Prior methods for model order selection which use MSWF-based
algorithms \cite{goldstein} or AVF-based recursions \cite{avf5} have
considered projection techniques \cite{goldstein} and
cross-validation \cite{avf5} approaches. Here, we focus on an
approach that jointly determines $D$ based on an LS criterion
computed by the estimators ${\boldsymbol S}_{D,j}[i]$ and
$\bar{\boldsymbol w}_j[i]$, where the superscript $D$ denotes the
rank used for the adaptation. The methods considered here (the
proposed and existing ones \cite{goldstein,avf5}) are the most
suitable for model-order adaptation in time-varying channels. Other
techniques such as the Akaike information criterion-based and the
minimum description length do not lend themselves to time-varying
situations and are {  computationally} complex \cite{haykin}.

The key quantities to be updated are the transformation matrix
${\boldsymbol S}_{D,j}[i]$, the reduced-rank estimator
$\bar{\boldsymbol w}_j[i]$, and the inverse of the reduced-rank
covariance matrix $\bar{\boldsymbol P}_j[i]$ (for the proposed RLS
algorithm). Specifically, we allow the dimensions of ${\boldsymbol
S}_{D,j}[i]$ and $\bar{\boldsymbol w}_j[i]$ to vary from $D_{\rm
min}$ and $D_{\rm max}$, which are the minimum and maximum ranks
allowed, respectively. It is important to note that only one
recursion to obtain $\bar{\boldsymbol P}_j[i]$ is computed with
$D_{\rm max}$ in order to keep the complexity low. Once
$\bar{\boldsymbol P}_j[i]$ is obtained, we perform a search for
the best $D$ for ${\boldsymbol S}_{D,j}[i]$ and $\bar{\boldsymbol
w}_j[i]$ that require sub-matrices of $\bar{\boldsymbol P}_j[i]$
for their computation. The transformation matrix $ {\boldsymbol
S}_{D,j}[i]$ and the reduced-rank estimator $\bar{\boldsymbol
w}_{j}[i]$ employed with this algorithm are illustrated by
\begin{equation}
\begin{split}
\hspace{-0.25em} {\boldsymbol S}_{D,j}[i] & =
\left[\hspace{-0.5em}\begin{array}{ccccc} s_{1,1,j}[i] & \ldots &
s_{1,D_{\rm min},j}[i] & \ldots & s_{1,D_{\rm max},j}[i]
\\ \vdots & \vdots & \vdots & \ddots & \vdots \\
s_{M,1,j}[i]  & \ldots & s_{M,D_{\rm min},j}[i] & \ldots &
s_{M,D_{\rm max},j}[i] \end{array} \hspace{-0.75em} \right] \\~~{\rm and}  ~~ \\
\bar{\boldsymbol w}_{D,j}[i] & = \left[\begin{array}{c} w_{1,j}[i]
~ w_{2,j}[i] ~ \ldots ~ w_{D_{\rm min},j}[i] ~ \ldots ~ w_{D_{\rm
max},j}[i]
\end{array}  \right]^T
\end{split}
\end{equation}
The method for automatically selecting $D$ of the algorithm is
based on the exponentially weighted \textit{a posteriori}
least-squares type cost function:
\begin{equation}
{\mathcal C}_j({\boldsymbol S}_{D,j}[i], {\bar{\boldsymbol
w}}_{d,j}[i]) = \sum_{l=1}^{i} \lambda^{i-l} \big| x_j[l] -
{\bar{\boldsymbol w}}^{H}_{d,j}[i]{{\boldsymbol S}}_{D,j}^{H}
[i]{\boldsymbol r}_j[l]|^2. \label{eq:costadap}
\end{equation}
For each time interval $i$, we select the rank $D_{\rm opt}[i]$
which minimizes ${\mathcal C}_j({\boldsymbol S}_{D,j}[i],
{\bar{\boldsymbol w}}_{D,j}[i])$ and the exponential weighting
factor $\lambda$ is required as the optimal rank varies as a
function of the data record. The transformation matrix $
{\boldsymbol S}_{D,j}[i]$ and the reduced-rank estimator
$\bar{\boldsymbol w}_{d,j}[i]$ are updated along with
$\bar{\boldsymbol P}[i]$ for the maximum allowed rank $D_{\rm
max}$ and then the proposed rank adaptation algorithm determines
the the best model order for each time instant $i$ using the cost
function in (\ref{eq:costadap}). The proposed model-order
selection algorithm is given by
\begin{equation}
D_{j,{\rm opt}}[i] = \arg \min_{D_{\rm min} \leq d \leq D_{\rm
max}} {\mathcal C}_j({\boldsymbol S}_{d,j}[i],\bar{\boldsymbol
w}_{d,j}[i]),
\end{equation}
where $d$ is an integer, $D_{\rm min}$ and $D_{\rm max}$ are the
minimum and maximum ranks allowed for the estimators, respectively.
A small rank may provide faster adaptation during the initial stages
of the estimation procedure, whereas a large rank usually yields a
better steady-state performance. Our studies indicate that the range
for which the rank $D$ of the proposed algorithms have a positive
impact on the performance of the algorithms is limited. {
Specifically, we have found that even for large systems
($N_R=N_T=20,30,40,50,60$) the rank does not scale with the system
size and remains small. The typical range of values remains between
$D_{\rm min}=3$ and $D_{\rm max}=8$ for the system sizes examined
($N_R=N_T=20,30,40,50,60$). This is an important aspect of the
proposed algorithms because it keeps the complexity low (comparable
to a standard RLS algorithm).} For the scenarios considered in what
follows, we set $D_{\rm min}=3$ and $D_{\rm max}=8$. In the
simulations section, we will illustrate how the proposed model-order
selection algorithm performs.

\subsection{Computational Complexity}

In this subsection, we illustrate the computational complexity
requirements of the proposed RLS algorithms and compare them with
those of existing algorithms. {  We also provide the computation
complexity of the proposed and existing model-order selection
algorithms.} { The computational complexity of the algorithms is
expressed} in terms of additions and multiplications, as depicted in
Table I. For the proposed reduced-rank RLS algorithm the complexity
is quadratic with $M=LN_R+B(N_T-1)$ and $D$. This amounts to a
complexity slightly higher than that observed for the full-rank RLS
algorithm, provided $D$ is significantly smaller than $M$, and
significantly less than the cost of the MSWF-RLS \cite{goldstein}
and the AVF \cite{avf5} algorithms. {  The complexity of the
proposed model-order selection algorithm is given in Table II.}

\begin{table}[t]
\centering%
\caption{\small Computational complexity of algorithms.
\vspace{-0.5em}} {
\begin{tabular}{lcc}
\hline
{\small Algorithm} & {\small Additions} & {\small Multiplications} \\
\emph{\small \bf Full-rank }\cite{haykin} & {\small
$2M^2 $} & {\small $3M^{2}$} \\
\emph{\small \bf   } & {\small $+M +1$} & {\small $+5M  $} \\
\hline
\emph{\small \bf Proposed}  & {\small $ 2M^2  $}  & {\small $ 3M^2  $}  \\
\emph{\small \bf }  & {\small $- M + 4D^2 $} & {\small $+3M+6D^2$}\\
\emph{\small \bf }  & {\small $+ MD + D+3$}  & {\small $+ MD +8D$}\\
\hline
\emph{\small \bf MSWF }\cite{goldstein} & {\small $DM^2$} & {\small $DM^2$} \\
\emph{\small \bf } & {\small $+6D^2-8D+2$} & {\small $+2DM+3D$} \\
\emph{\small \bf } & {\small $+M^2$} & {\small $+M^2+2$} \\
\hline
\emph{\small \bf AVF }\cite{avf5} & {\small $DM^2+2M-1$} & {\small $4DM^2$} \\
\emph{\small \bf } & {\small $+5D(M-1)+1$} & {\small $+4DM+4M $} \\
\emph{\small  } & {\small $+3(DM-1)^2$} & {\small $+ 4D+ 2$} \\
\hline
\end{tabular}
}
\end{table}

\begin{figure}[!htb]
\begin{center}
\def\epsfsize#1#2{1\columnwidth}
\epsfbox{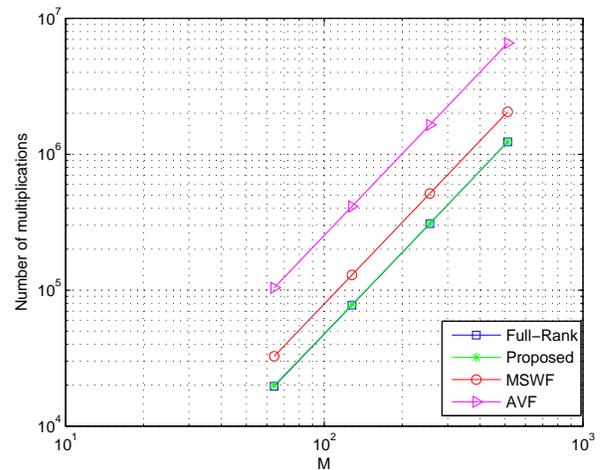}  \vspace{-0.5em} \caption{\small Complexity in
terms of multiplications against number of input samples ($M$) with
$D=5$, $N_T=N_R$, $L=8$, $B=2$ .} \label{comp}
\end{center}
\end{figure}

In order to illustrate the main trends and requirements in terms
of complexity of the proposed and existing algorithms, we show in
Fig. \ref{comp} the complexity against the number of input samples
$M$ for the parameters $D=5$, $N_T=N_R$, $L=8$ and $B=2$. The
curves indicate that the proposed reduced-rank RLS algorithm has a
complexity significantly lower than the MSWF-RLS algorithm
\cite{goldstein} and the AVF \cite{avf5}, whereas it remains at
the same level of the full-rank RLS algorithm.

{  The computational complexity of the model-order selection
algorithms including the proposed and the existing techniques is
shown in Table II. We can notice that the proposed model-order
selection algorithm is significantly less complex than the existing
methods based on projection with stopping rule \cite{goldstein} and
the CV approach \cite{avf5}. Specifically, the proposed algorithm
that uses extended filters only requires $2(D_{\rm max}-D_{\rm
min})$ additions, as depicted in the first row of Table II. To this
cost we must add the operations required by the proposed RLS
algorithm, whose complexity is shown in the second row of Table I
using $D_{\rm max}$. The complexities of the MSWF and the AVF
algorithms are detailed in the third and fourth rows of Table I. For
their operation with model-order selection algorithms, a designer
must add their complexities in Table I to the complexity of the
model-order selection algorithms of interest in Table II.}

\begin{table}[t]
\centering%
\caption{\small Computational complexity of model-order
selection algorithms.} {
\begin{tabular}{lcc}
\hline
{\small Algorithm} & {\small Additions} & {\small Multiplications} \\
\emph{\small  Proposed }  & {\small $ 2(D_{\rm max}- D_{\rm min}) +1$}  & {\small $ -$}  \\
\hline
\emph{\small  Projection with } & {\small
$2(2M-1)\times $} & {\small $(M^2+M+1)\times $} \\
\emph{\small  Stopping Rule }\cite{goldstein} & {\small
$(D_{\rm max}- D_{\rm min})+1$} & {\small $(D_{\rm max}-D_{\rm
min}+1)$} \\
\hline
\emph{\small  CV }\cite{avf5} & {\small $(2M-1)\times $} & {\small $(D_{\rm max}- D_{\rm min}+1)\times$} \\
\emph{\small \bf  } & {\small $(2(D_{\rm max}- D_{\rm min}) +1)$}
& {\small $M + 1$}
\\
\hline
\end{tabular}
}
\end{table}

\section{Analysis of The Proposed Algorithms}

In this section, we conduct an analysis of the proposed algorithms
that compute the estimators ${\boldsymbol S}_{D,j}[i]$ and
${\bar{\boldsymbol w}}_j[i]$ of the proposed scheme. We first
highlight the alternating optimization nature of the proposed
algorithms and make use of recent convergence results for this class
of algorithms \cite{csiszar,niesen}. In particular, we present a set
of sufficient conditions under which the proposed algorithms
converge to the optimal estimators. This is corroborated by our
numerical studies that verify that the method is insensitive to
different initializations (except for the case when ${\boldsymbol
S}_{D,j}[i]$ is a null matrix which annihilates the received signal)
and that it converges to the same point of minimum. We establish the
global convergence of the proposed algorithm via induction and show
that that the sequence of estimators ${\boldsymbol S}_{D,j}[i]$ and
${\bar{\boldsymbol w}}_j[i]$ produces a sequence of outputs that is
bounded and converges to the reduced-rank Wiener filter
\cite{scharfo},\cite{hua1}.

\subsection{Sufficient Conditions for Convergence}

In order to develop the analysis and proofs, we need to define a
metric space and the Hausdorff distance that will be used
extensively. A metric space is an ordered pair $({\mathcal M}, d)$
where ${\mathcal M}$ is a non-empty set and $d$ is a metric on
${\mathcal M}$, i.e., a function $d: {\mathcal M} \times {\mathcal
M} \rightarrow {\mathbb R}$ such that for any $x$, $y$, $z$, and ${\mathcal M}$ we have:\\
a) $d(x,y) \geq 0$.\\
b) $d(x,y) =0$ $iff$ $x=y$.\\
c) $d(x,y) = d(y,x)$.\\
d) $d(x,z) \leq d(x,y) + d(y,z$  (triangle inequality).\\
The Hausdorff distance measures how far two subsets of a metric
space are from each other and is defined by
\begin{equation}
d_H(X,Y) = \max  \{ \sup_{x \in X}  \inf_{y \in Y} d(x,y), \sup_{y
\in Y}  \inf_{x \in X} d(x,y)  \}
\end{equation}
The proposed LS and RLS algorithms can be stated as an alternating
minimization strategy based on the sum of error squares (SES)
defined in (\ref{ses}) and expressed as
\begin{equation}
{\boldsymbol S}_{D,j}[i]  \in  \arg \min_{{\boldsymbol
S}_{D,j}^{\rm opt} \in \underline{\boldsymbol S}_{D,j}[i]}  {\rm SES}
({\boldsymbol S}_{D,j}^{\rm opt}, \bar{\boldsymbol w}_j[i])
\end{equation}
\begin{equation}
\bar{\boldsymbol w}_j[i]  \in  \arg \min_{\bar{\boldsymbol
w}_{j}^{\rm opt} \in \underline{\bar{\boldsymbol w}}_j[i]}  {\rm SES}
({\boldsymbol S}_{D,j}[i], \bar{\boldsymbol w}^{\rm opt}_j),
\end{equation}
{ where ${\boldsymbol S}_{D,j}^{\rm opt}$ and $\bar{\boldsymbol
w}^{\rm opt}_j$ correspond to the optimal values of ${\boldsymbol
S}_{D,j}[i]$ and $\bar{\boldsymbol w}_j[i]$, respectively, and the
sequences of compact sets $\{ \underline{\boldsymbol S}_{D,j}[i]
\}_{i \geq 0}$ and $\{ \underline{\bar{\boldsymbol w}}_j[i] \}_{i
\geq 0}$ converge to the sets $ \underline{\boldsymbol S}_{D,{\rm
opt}}$ and $\underline{\bar{\boldsymbol w}}_{j, {\rm opt}}$,
respectively.}

{ Although we are not given the sets $ \underline{\boldsymbol
S}_{D,{\rm opt}}$ and $\underline{\bar{\boldsymbol w}}_{j, {\rm
opt}}$ directly, we observe the sequence of compact sets $\{
\underline{\boldsymbol S}_{D,j}[i] \}_{i \geq 0}$ and $\{
\underline{\bar{\boldsymbol w}}_j[i] \}_{i \geq 0}$. The goal of the
proposed algorithms is to find a sequence of ${\boldsymbol
S}_{D,j}[i]$ and $\bar{\boldsymbol w}_j[i]$ such that}
\begin{equation}
\lim_{i \rightarrow \infty} {\rm SES} ({\boldsymbol S}_{D,j}[i],
\bar{\boldsymbol w}_j[i]) = {\rm SES} ({\boldsymbol S}_{D,j}^{\rm
opt}, \bar{\boldsymbol w}^{\rm opt}_j)
\end{equation}
In order to present a set of sufficient conditions under which the
proposed algorithms converge, we need to so-called "three-point"
and "four-point" properties \cite{csiszar,niesen}. Let us assume
that there is a function $f: {\mathcal M} \times {\mathcal M}
\rightarrow {\mathbb R}$ such that the following conditions are
satisfied:\\

{ ${\mathbf 1}$) Three-point property (${\boldsymbol S}_{D,j}^{\rm
opt}$, $\tilde{\boldsymbol S}_{D,j}$,  $\bar{\boldsymbol w}_j^{\rm
opt}$): for all $i \geq 1$, ${\boldsymbol S}_{D,j}^{\rm opt} \in
\underline{\boldsymbol S}_{D,j}[i]$, $\bar{\boldsymbol w}_j^{\rm
opt} \in \underline{\bar{\boldsymbol w}}_j[i]$, and \\
$\tilde{\boldsymbol S}_{D,j} \in \arg \min_{ \bar{\boldsymbol
w}_j^{\rm opt} \in \underline{\bar{\boldsymbol w}}_j[i]} {\rm SES}
({\boldsymbol S}_{D,j}^{\rm opt},
\bar{\boldsymbol w}_j^{\rm opt} )$\\
\begin{equation}
f( {\boldsymbol S}_{D,j}^{\rm opt},\tilde{\boldsymbol S}_{D,j}) +
{\rm SES} (\tilde{\boldsymbol S}_{D,j},  \bar{\boldsymbol
w}_j^{\rm opt}) \leq {\rm SES} ({\boldsymbol S}_{D,j}^{\rm opt},
\bar{\boldsymbol w}_j^{\rm opt}). \label{cond1}
\end{equation}

${\mathbf 2}$) Four-point property (${\boldsymbol S}_{D,j}^{\rm
opt}$,  $\bar{\boldsymbol w}_j^{\rm opt}$, $\tilde{\boldsymbol
S}_{D,j}$, $\tilde{\bar{\boldsymbol w}}_j^{\rm opt}$): for all $i
\geq 1$, ${\boldsymbol S}_{D,j}^{\rm opt}$, $\tilde{\boldsymbol
S}_{D,j} \in \underline{\boldsymbol S}_{D,j}[i]$, $\bar{\boldsymbol
w}_j^{\rm opt} \in \underline{\bar{\boldsymbol w}}_j[i]$, and
$\tilde{\bar{\boldsymbol w}}_{D,j} \in \arg \min_{
\bar{\boldsymbol w}_j^{\rm opt} \in \underline{\bar{\boldsymbol w}}_j[i]} {\rm
SES} (\tilde{\boldsymbol S}_{D,j}, \bar{\boldsymbol w}_j^{\rm
opt} )$ \\
\begin{equation}
{\rm SES} ({\boldsymbol S}_{D,j}^{\rm opt},
\tilde{\bar{\boldsymbol w}}_j) \leq {\rm SES} ({\boldsymbol
S}_{D,j}^{\rm opt}, \bar{\boldsymbol w}_j^{\rm opt}) +
f({\boldsymbol S}_{D,j}^{\rm opt},\tilde{\boldsymbol S}_{D,j}).
\label{cond2}
\end{equation}

{\it Theorem}: Let $\{ (\underline{\boldsymbol S}_{D,j}[i], \underline{\bar{\boldsymbol
w}}_j[i]) \}_{i \geq 0}$, $\underline{\boldsymbol S}_{D,j}^{\rm opt}$,
$\underline{\bar{\boldsymbol w}}_j^{\rm opt}$ be compact subsets of the compact
metric space $({\mathcal M}, d)$ such that
\begin{equation}
\underline{\boldsymbol S}_{D,j}[i] \xrightarrow{d_H} \underline{\boldsymbol S}_{D,j}^{\rm
opt}, ~~~~ \underline{\bar{\boldsymbol w}}_j[i] \xrightarrow{d_H} \underline{\bar{\boldsymbol
w}}_j^{\rm opt}
\end{equation}
and let ${\rm SES}: {\mathcal M} \times {\mathcal M} \rightarrow
{\mathbb R}$ be a continuous function.}

Now let conditions ${\mathbf 1}$) and ${\mathbf 2}$) hold. Then,
for the proposed algorithms we have
\begin{equation}
\lim_{i \rightarrow \infty} {\rm SES} ({\boldsymbol S}_{D,j}[i],
\bar{\boldsymbol w}_j[i]) = {\rm SES} ({\boldsymbol S}_{D,j}^{\rm opt}
, \bar{\boldsymbol w}_j^{\rm opt})
\end{equation}
A general proof of this theorem is detailed in
\cite{csiszar,niesen}.

\subsection{Convergence to the Optimal Reduced-Rank Estimator}

In this subsection, we show that the proposed reduced-rank algorithm
converges globally and exponentially to the optimal reduced-rank
estimator \cite{scharfo},\cite{hua1}. {  We assume that $1 \geq
\lambda \gg 0$ (equal or close to one) , the desired product of the
optimal solutions}, i.e., ${\boldsymbol w}_{j}^{\rm opt} =
{\boldsymbol S}_{D,j}^{\rm opt} \bar{\boldsymbol w}_{j}^{\rm opt}$
is known and given by ${\boldsymbol R}^{-1/2}_j[i] \big(
{\boldsymbol R}^{-1/2}_j[i]{\boldsymbol p}_j[i] \big)_{1:D}$
\cite{haykin},\cite{hua1}, where ${\boldsymbol R}^{-1/2}_j[i]$ is
the square root of the input data covariance matrix and the
subscript ${1:D}$ denotes truncation of the subspace.

In order to proceed with our proof, let us rewrite the expressions
in (\ref{filtersd}) and (\ref{filterw}) for time instant $0$ as
follows
\begin{equation}
{\boldsymbol R}_j[0]{\boldsymbol S}_{D,j}[0] {\boldsymbol
R}_{{\bar w}_j}[0] = {\boldsymbol P}_{D,j}[0] = {\boldsymbol
p}_j[0]{\bar{\boldsymbol w}}_j^H[0] , \label{filttalt}
\end{equation}
\begin{equation}
\bar{\boldsymbol R}_j[0]{\bar{\boldsymbol w}}_j[1] = {\boldsymbol
S}_{D,j}^H[0]{\boldsymbol R}_j[0]{\boldsymbol
S}_{D,j}[0]{\bar{\boldsymbol w}}_j[1] = \bar{\boldsymbol p}_j[0],
\label{filtwalt}
\end{equation}
Using (\ref{filttalt}) we can obtain the following relation
\begin{equation}
{\boldsymbol R}_{{\bar w}_j}[0] = \big({\boldsymbol
S}_{D,j}^H[0]{\boldsymbol R}^2_j[0]{\boldsymbol
S}_{D,j}[0]\big)^{-1} {\boldsymbol S}_{D,j}^H[0]{\boldsymbol
R}_j[0]{\boldsymbol p}_j[0]{\bar{\boldsymbol w}}_j^H[0] ,
\label{filtrw}
\end{equation}
Substituting the above result for ${\boldsymbol R}_{w_j}[0]$ into
the expression in (\ref{filttalt}) we get a recursive expression
for ${\boldsymbol S}_{D,j}[0]$
\begin{equation}
\begin{split}
{\boldsymbol S}_{D,j}[0] & = {\boldsymbol R}_j[0]^{-1}
{\boldsymbol p}_j[0]{\bar{\boldsymbol w}}_j^H[0] \big(
{\boldsymbol S}_{D,j}^H[0]{\boldsymbol R}_j[0]{\boldsymbol
p}_j[0]{\bar{\boldsymbol w}}_j^H[0]\big)^{-1} \cdot \\ &  \quad
\cdot \big({\boldsymbol S}_{D,j}^H[0]{\boldsymbol
R}^2_j[0]{\boldsymbol S}_{D,j}[0]\big)^{-1}, \label{filtrwf}
\end{split}
\end{equation}
Using (\ref{filtwalt}) we can express $\bar{\boldsymbol w}_j[1]$
as
\begin{equation}
{\bar{\boldsymbol w}}_j[1] = \big({\boldsymbol
S}_{D,j}^H[0]{\boldsymbol R}_j[0]{\boldsymbol S}_{D,j}[0]
\big)^{-1} {\boldsymbol S}_{D,j}^H[0]{\boldsymbol p}_j[0],
\label{filtwinst}
\end{equation}
Employing the relation ${\boldsymbol w}_j[1] = {\boldsymbol
S}_{D,j}[1] \bar{\boldsymbol w}_j[1]$, we obtain
\begin{equation}
\begin{split}
{{\boldsymbol w}}_j[1] & =  {\boldsymbol R}_j[1]^{-1} {\boldsymbol
p}[1]{\bar{\boldsymbol w}}_j^H[1] \big( {\boldsymbol
S}_{D,j}^H[1]{\boldsymbol R}_j[1]{\boldsymbol
p}_j[1]{\bar{\boldsymbol w}}_j^H[1]\big)^{-1} \cdot
\\ &  \quad  \cdot \big({\boldsymbol S}_{D,j}^H[1]{\boldsymbol
R}^2_j[1]{\boldsymbol S}_{D,j}[1]\big)^{-1}\big({\boldsymbol
S}_{D,j}^H[0]{\boldsymbol R}_j[0]{\boldsymbol S}_{D,j}[0]
\big)^{-1} {\boldsymbol S}_{D,j}^H[0]{\boldsymbol p}_j[0]
\end{split}
\end{equation}
More generally, we can express the proposed reduced-rank LS
algorithm by the following recursion
\begin{equation}
\begin{split}
{{\boldsymbol w}}_j[i] & = {\boldsymbol S}_{D,j}[i]
\bar{\boldsymbol w}_j[i] \\ & =  {\boldsymbol R}_j[i]^{-1}
{\boldsymbol p}_j[i]{\bar{\boldsymbol w}}_j^H[i] \big(
{\boldsymbol S}_{D,j}^H[i]{\boldsymbol R}_j[i]{\boldsymbol
p}_j[i]{\bar{\boldsymbol w}}_j^H[i]\big)^{-1} \cdot
\\ &  \quad  \cdot \big({\boldsymbol S}_{D,j}^H[i]{\boldsymbol
R}^2[i]{\boldsymbol S}_{D,j}[i]\big)^{-1} \cdot
\\ &  \quad  \cdot \big({\boldsymbol S}_{D,j}^H[i-1]{\boldsymbol R}_j[i-1]{\boldsymbol S}_{D,j}[i-1]
\big)^{-1} {\boldsymbol S}_{D,j}^H[i-1]{\boldsymbol
p}_j[i-1]\label{filtwconc}.
\end{split}
\end{equation}
Since the optimal reduced-rank filter can be described by the EVD of
${\boldsymbol R}^{-1/2}_j[i]{\boldsymbol p}_j[i]$ \cite{scharfo},
\cite{hua1}, where ${\boldsymbol R}^{-1/2}_j[i]$ is the square root
of the covariance matrix ${\boldsymbol R}_j[i]$ and ${\boldsymbol
p}_j[i]$ is the cross-correlation vector, then we have
\begin{equation}
{\boldsymbol R}^{-1/2}_j[i]{\boldsymbol p}_j[i] = {\boldsymbol
\Phi}_j {\boldsymbol \Lambda}_j {\boldsymbol \Phi}^H_j {\boldsymbol
p}_j[i], \label{optfilt}
\end{equation}
where ${\boldsymbol \Lambda}_j$ is a $M\times M$ diagonal matrix
with the eigenvalues of ${\boldsymbol R}_j$ and ${\boldsymbol
\Phi}_j$ is a $M\times M$ unitary matrix with the eigenvectors of
${\boldsymbol R}_j$. Assuming that there exists some ${\boldsymbol
w}_j[0]$ such that the randomly selected ${\boldsymbol S}_{D,j}[0]$
can be written as \cite{hua1}
\begin{equation}
{\boldsymbol S}_{D,j}[0] = {\boldsymbol R}^{-1/2}_j[i]{\boldsymbol
\Phi}_j {\boldsymbol w}_j[0]. \label{tdo}
\end{equation}
Using (\ref{tdo}) and (\ref{optfilt}) in (\ref{filtwconc}), and
manipulating the algebraic expressions, we can express
(\ref{filtwconc}) in a more compact way that is suitable for
analysis, as given by
\begin{equation}
{\boldsymbol w}_j[i] = {\boldsymbol \Lambda}^2_j {\boldsymbol
w}_j[i-1] ({\boldsymbol w}^H_j[i-1] {\boldsymbol \Lambda}^2_j
{\boldsymbol w}_j[i-1] )^{-1} {\boldsymbol w}^H_j[i-1]
{\boldsymbol w}_j[i-1]. \label{wdecomp}
\end{equation}
The above expression can be decomposed as follows
\begin{equation}
{\boldsymbol w}_j[i] = {\boldsymbol Q}_j[i]~{\boldsymbol
Q}_j[i-1]~\ldots~{\boldsymbol Q}_j[1]~{\boldsymbol w}_j[0],
\label{wdec}
\end{equation}
where
\begin{equation}
{\boldsymbol Q}_j[i] = {\boldsymbol \Lambda}^{2i}_j {\boldsymbol
w}_j[0]( {\boldsymbol w}^H_j[0] {\boldsymbol \Lambda}^{4i-2}_j
{\boldsymbol w}_j[0])^{-1} {\boldsymbol w}^H_j[0] {\boldsymbol
\Lambda}^{2i-2}_j \label{qexp}.
\end{equation}
At this point, we need to establish that the norm of ${\boldsymbol
S}_{D,j}[i]$ for all $i$ is both lower and upper bounded, i.e.,
$0< ||{\boldsymbol S}_{D,j}[i]|| < \infty $ for all $i$, and that
${\boldsymbol w}_j[i] = {\boldsymbol S}_{D,j}[i] \bar{\boldsymbol
w}_j[i]$ approaches ${\boldsymbol w}_{j,{\rm opt}}[i]$
exponentially as $i$ increases. Due to the linear mapping, the
boundedness of ${\boldsymbol S}_{D,j}[i]$ is equivalent to that of
${\boldsymbol w}_j[i]$. Therefore, we have upon convergence
${\boldsymbol w}^H_j[i]{\boldsymbol w}_j[i-1] = {\boldsymbol
w}^H_j[i-1] {\boldsymbol w}_j[i-1]$. Since $||{\boldsymbol
w}^H_j[i]{\boldsymbol w}_j[i-1]|| \leq || {\boldsymbol w}_j[i-1]||
|| {\boldsymbol w}_j[i]||$ and $||{\boldsymbol
w}^H_j[i-1]{\boldsymbol w}_j[i-1]|| = || {\boldsymbol
w}_j[i-1]||^2$, the relation ${\boldsymbol w}^H_j[i]{\boldsymbol
w}_j[i-1] = {\boldsymbol w}^H_j[i-1] {\boldsymbol w}_j[i-1]$
implies $||{\boldsymbol w}_j[i] || > ||{\boldsymbol w}_j[i-1]||$
and hence
\begin{equation}
||{\boldsymbol w}_j[\infty]|| \geq || {\boldsymbol w}_j[i]|| \geq
||{\boldsymbol w}_j[0]||
\end{equation}
In order to show that the upper bound $||{\boldsymbol
w}_j[\infty]||$ is finite, let us express the $M \times M$ matrix
${\boldsymbol Q}_j[i]$ as a function of the $M \times 1$ vector
${\boldsymbol w}_j[i] = \left[
\begin{array}{c} {\boldsymbol w}_{j,1}[i] \\ {\boldsymbol w}_{j,2}[i] \end{array} \right]$
and the $M \times M$ matrix $ {\boldsymbol \Lambda} = \left[
\begin{array}{cc} {\boldsymbol
\Lambda}_{j,1} & \\
& {\boldsymbol \Lambda}_{j,2} \end{array} \right]$. Substituting
the previous expressions of ${\boldsymbol w}_j[i]$ and
${\boldsymbol \Lambda}_j$ into ${\boldsymbol Q}_j[i]$ given in
(\ref{qexp}), we obtain
\begin{equation}
\begin{split}
{\boldsymbol Q}_j[i] & = \left[
\begin{array}{c} {\boldsymbol
\Lambda}_{j,1}^{2i} {\boldsymbol w}_{j,1}[0] \\ {\boldsymbol
\Lambda}_{j,2}^{2i} {\boldsymbol w}_{j,2}[0] \end{array} \right] (
{\boldsymbol w}^H_{j,1}[0] {\boldsymbol \Lambda}^{4i-2}_{j,1}
{\boldsymbol w}_{j,1}[0] \\ & \quad + {\boldsymbol w}^H_{j,2}[0]
{\boldsymbol \Lambda}^{4i-2}_2 {\boldsymbol w}_{j,2}[0])^{-1}
\left[
\begin{array}{c} {\boldsymbol w}_{j,1}^H[0]{\boldsymbol
\Lambda}_{j,1}^{2i-2}  \\  {\boldsymbol w}_{j,2}^H[0]{\boldsymbol
\Lambda}_{j,2}^{2i-2} \end{array} \right] \label{qexp2}.
\end{split}
\end{equation}
Using the matrix identity $({\boldsymbol A} + {\boldsymbol
B})^{-1} = {\boldsymbol A}^{-1} - {\boldsymbol A}^{-1}{\boldsymbol
B} ({\boldsymbol I} + {\boldsymbol A}^{-1}{\boldsymbol B})^{-1}
{\boldsymbol A}^{-1}$ to the decomposed ${\boldsymbol Q}_j[i]$ in
(\ref{qexp2}) and making $i$ large, we get
\begin{equation}
{\boldsymbol Q}_j[i] = {\rm diag} \big( \underbrace{1 \ldots
1}_{D}~ \underbrace{0 \ldots 0}_{M-D} \big) + {\rm
O}(\epsilon[i]). \label{Qres}
\end{equation}
where $\epsilon[i] = (\lambda_{r+1}/ \lambda_{r})^{2i}$ with
$\lambda_{r+1}$ and $\lambda_r$ are the $(r+1)$th, the $r$th largest
singular values of ${\boldsymbol R}^{-1/2}_j[i]{\boldsymbol p}_j[i]$
and ${\rm O}(\cdot)$ denotes the order of the argument. From
(\ref{Qres}), it follows that for some positive constant $k$, we
have $||{\boldsymbol w}_j[i]|| \leq 1 + k \epsilon[i]$. From
(\ref{wdec}), we obtain
\begin{equation}
\begin{split}
||{\boldsymbol w}_j[\infty]|| & \leq ||{\boldsymbol Q}_j[\infty]||
\ldots ||{\boldsymbol Q}_j[2]||~||{\boldsymbol Q}_j[1]||~ ||
{\boldsymbol Q}_j[0]|| \\ & \leq ||{\boldsymbol w}_j[0]||
\prod_{i=1}^{\infty} (1+ k \epsilon[i]) \\
& = ||{\boldsymbol w}_j[0]|| \exp \Big(\sum_{i=1}^{\infty} log (1+
k
\epsilon[i])\Big) \\
& \leq ||{\boldsymbol w}_j[0]|| \exp \Big( \sum_{i=1}^{\infty} k
\epsilon[i] \Big) \\ &  = ||{\boldsymbol w}_j[0]|| \exp \Big(
\frac{k}{1- (\lambda_{r+1}/\lambda_r)^2 } \Big)
\end{split}
\end{equation}
With the development above, the norm of ${\boldsymbol w}_j[i]$ is
proven to be both lower and upper bounded. Once this is established,
the expression in (\ref{filtwconc}) converges for a sufficiently
large $i$ to the reduced-rank Wiener filter. This is verified by
equating the terms of (\ref{wdecomp}) which yields
\begin{equation}
\begin{split}
{{\boldsymbol w}}_j[i] & =  {\boldsymbol R}_j[i]^{-1} {\boldsymbol
p}_j[i]{\bar{\boldsymbol w}_j}^H[i] \big( {\boldsymbol
S}_{D,j}^H[i]{\boldsymbol R}_j[i]{\boldsymbol
p}_j[i]{\bar{\boldsymbol w}_j}^H[i]\big)^{-1} \cdot
\\ &  \quad \cdot \big({\boldsymbol
S}_{D,j}^H[i]{\boldsymbol R}^2_j[i]{\boldsymbol
S}_{D,j}[i]\big)^{-1} \cdot
\\ &  \quad  \cdot \big({\boldsymbol S}_{D,j}^H[i-1]{\boldsymbol R}_j[i-1]{\boldsymbol S}_{D,j}[i-1]
\big)^{-1} {\boldsymbol S}_{D,j}^H[i-1]{\boldsymbol p}_j[i-1] \\
&  = {\boldsymbol R}^{-1/2}_j[i] {\boldsymbol \Phi}_{j,1}
{\boldsymbol \Lambda}_{j,1} {\boldsymbol \Phi}_{j,1}^{H}{\boldsymbol
p}_j[i] + {\rm O}(\epsilon[i]) \label{filtwconc2},
\end{split}
\end{equation}
where ${\boldsymbol \Phi}_1$ is a $M \times D$ matrix with the $D$
largest eigenvectors of ${\boldsymbol R}_j[i]$ and ${\boldsymbol
\Lambda}_{j,1}$ is a $D \times D$ matrix with the largest
eigenvalues of ${\boldsymbol R}_j[i]$.

\section{Simulation Results}

In this section, we evaluate the bit error rate~(BER) performance of
the proposed MIMO equalization structure, algorithms and existing
techniques, namely, the full-rank \cite{dhahir}, the reduced-rank
MSWF \cite{goldstein}, and the AVF \cite{avf5} techniques for the
design of the receivers. For all simulations and the proposed
reduced-rank RLS algorithm, we use the initial values
$\bar{\boldsymbol w}_j[0]= \big[ 1, 0, \ldots, 0 \big]$ and
${\boldsymbol S}_{D,j}[0]=[{\boldsymbol I}_D ~ {\boldsymbol 0}_{D
\times (M-D)}]^T$. For the next experiments, we adopt an observation
window of $L=8$, the multipath channels ({ the channel vectors
${\boldsymbol h}_{j,k}[i] = [h_{j,k,1}[i] ~ \ldots ~
h_{j,k,L_p-1}]^T$}) are modelled by FIR filters with $L_p$
coefficients spaced by one symbol and the system employs QPSK
modulation. The channel is time-varying over the transmitted
packets, the profile follows the UMTS Vehicular A channel model
\cite{umts} with $L_p=5$ and the fading is given by Clarke's model
\cite{rappa}. We average the experiments over $1000$ runs and define
the signal-to-noise ratio (SNR) as $\textrm{SNR} = 10 \log_{10}
\frac{N_T \sigma_x^2}{\sigma^2}$, where $\sigma_x^2$ is the variance
of the transmitted symbols and $\sigma^2$ is the noise variance. The
adaptive MIMO equalizers employ $N_T=4$, $B=4$, $L=8$ and $N_R=8$ in
a spatial multiplexing configuration, leading to estimators at the
receiver with $M=LN_R + B(N_T-1)=72$ coefficients. The adaptive RLS
estimators of all methods are trained with $250$ symbols, {  employ
$\lambda=0.998$ unless otherwise specified}, and are then switched
to decision-directed mode.

\subsection{Convergence Performance and Impact of Rank}

In the first experiment, we consider the BER performance versus the
rank $D$ with optimized parameters (forgetting factor
$\lambda=0.998$) for linear MIMO equalizers. The curves in Fig. 4
show that the best rank for the proposed scheme is $D=4$, which is
the closest among the analyzed algorithms to the optimal linear MMSE
that assumes the knowledge of the channel and the noise variance. In
addition, it should be remarked that our studies with systems with
different sizes show that the optimal rank $D$ does not vary
significantly with the system size. It remains in a small range of
values, which brings considerably faster convergence speed. However,
It should also be remarked that the optimal rank $D$ depends on the
data record size and other parameters of the systems.

\begin{figure}[!htb]
\begin{center}
\def\epsfsize#1#2{1\columnwidth}
\epsfbox{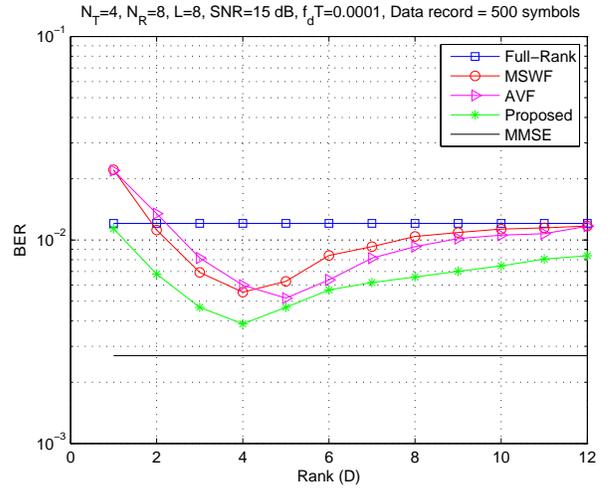} \caption{BER performance versus rank (D) for
linear MIMO equalizers.}
\end{center}
\end{figure}

\begin{figure}[!htb]
\begin{center}
\def\epsfsize#1#2{1\columnwidth}
\epsfbox{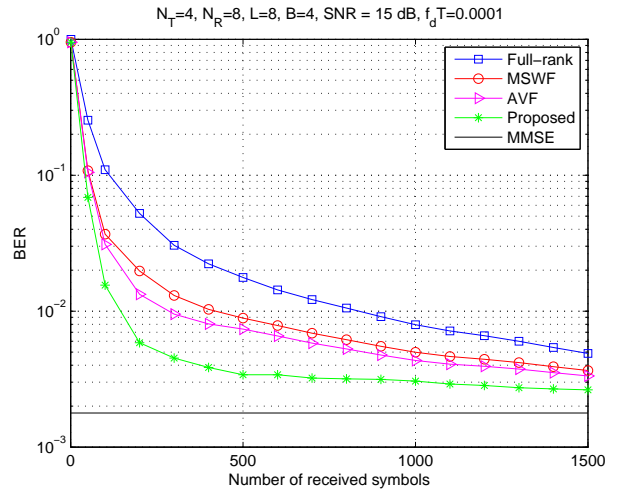} \caption{BER performance versus number of
received symbols. }
\end{center}
\end{figure}

The BER convergence performance versus the number of received
symbols for MIMO decision feedback equalizers with optimized but
fixed ranks is shown in Fig. 5. The results show that the proposed
scheme has a significantly faster convergence performance and
obtains good gains over the best known schemes. The plots show that
the proposed reduced-rank MIMO equalizer extends the dimensionality
reduction and its benefits such as fast convergence and robustness
against errors to the MIMO equalization task. The proposed RLS
estimation algorithm has the best performance and is followed by the
AVF, the MSWF, and the full-rank estimators. {  Note that the BER of
the considered techniques will converge to the same values if the
number of received symbols is very large and the channel is static.}

\subsection{Performance with Model Order Selection}

As previously mentioned in Section IV, it is possible to further
increase the convergence speed and { enhance} the tracking
performance of the reduced-rank algorithms using an automatic
model-order selection algorithm. In the next experiment, we consider
the proposed reduced-rank structures and algorithms with linear and
DF equalizers and compare their performance with fixed ranks and the
proposed automatic model-order selection algorithm developed in
Section IV.C. The results illustrated in Fig. 6 show that the
proposed model-order selection algorithm can effectively speed up
the convergence of the proposed reduced-rank RLS algorithm and {
ensure that it obtains an excellent } tracking performance. In what
follows, we will consider the proposed model-order selection
algorithm in conjunction with the proposed reduced-rank RLS
algorithm, and for a fair comparison we will equip the MSWF and the
AVF algorithms with the rank adaptation techniques reported in
\cite{goldstein} and \cite{avf5}, respectively.

\begin{figure}[!htb]
\begin{center}
\def\epsfsize#1#2{1\columnwidth}
\epsfbox{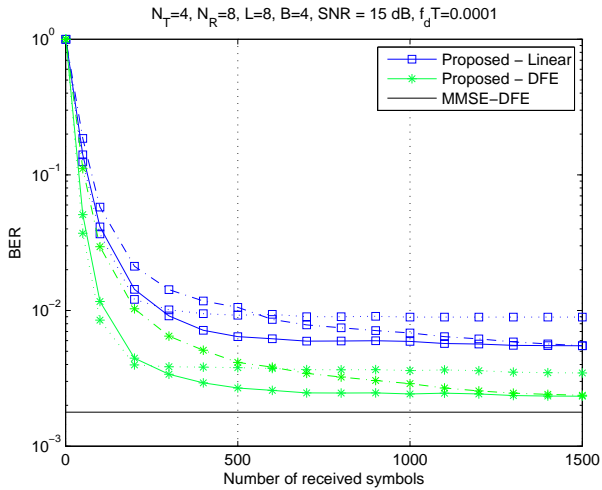} \caption{BER performance versus number of
received symbols for proposed estimation algorithms and structures.
The performance of the proposed reduced-rank algorithms is shown for
the proposed model-order selection algorithm (solid lines), for
$D=3$ (dotted lines) and for $D=8$ (dash-dotted lines).}
\end{center}
\end{figure}

\subsection{Performance for Various SNR and $f_dT$ Values}

The BER performance versus the signal-to-noise ratio (SNR) for MIMO
decision feedback equalizers operating with the automatic
model-order selection algorithms is shown in Fig. 7. The curves show
a significant advantage of reduced-rank algorithms over the
full-rank RLS algorithm. Specifically, the reduced-rank AVF and MSWF
techniques obtain gains of up to $3$ dB in SNR for the same BER over
the full-rank algorithm, whereas the proposed reduced-rank RLS
algorithm achieves a gain of up to $3$ dB over the AVF, the second
best reduced-rank algorithm. {  The main reasons for the differences
in diversity order are the speed and the level of accuracy of the
parameter estimation of the proposed and existing methods. If we
increase the number of received symbols to a very large value then
the diversity order attained by the different analyzed algorithms
would be the same as verified in our studies.}

\begin{figure}[!htb]
\begin{center}
\def\epsfsize#1#2{1\columnwidth}
\epsfbox{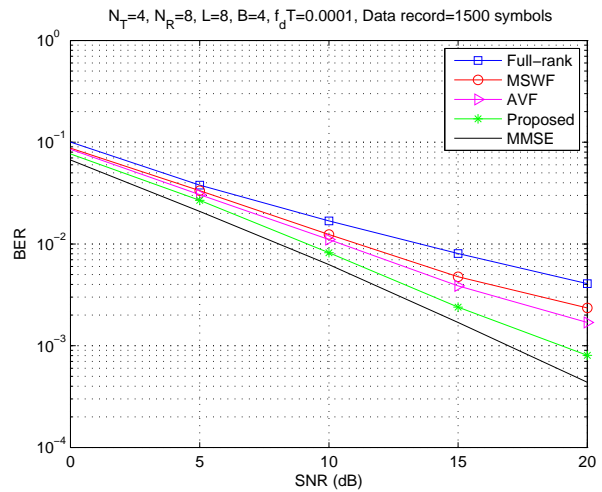} \caption{BER performance versus $SNR$ .}
\end{center}
\end{figure}

In order to assess the performance of the reduced-rank algorithms
for different fading rates, we consider an experiment where we
measure the BER of the proposed and analyzed algorithms against the
normalized fading rate $f_dT$ in cycles per symbol, where $f_d$ is
the maximum Doppler frequency and $T$ is the symbol rate. It should
be noted that the forgetting factor $\lambda$ was optimized for each
value of $f_dT$ { in} this experiment. {  In practice, a designer
could employ a mechanism to automatically adjust $\lambda$.} The
results of this experiment are shown in Fig. 8, where the advantages
of the reduced-rank algorithms and their superior performance in
time-varying scenarios is verified again.

\begin{figure}[!htb]
\begin{center}
\def\epsfsize#1#2{1\columnwidth}
\epsfbox{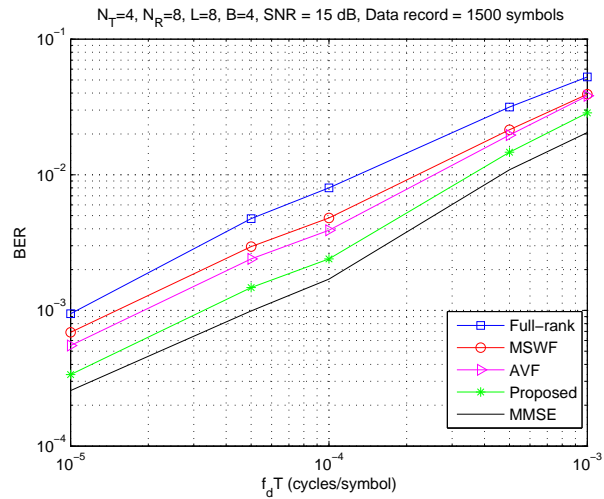} \caption{BER performance versus the normalized
fading rate $f_dT$ .}
\end{center}
\end{figure}

\subsection{Performance for MIMO-OFDM Systems}

In the previous experiments, we considered the proposed MIMO
equalization structure and algorithms for time-varying channels
that dynamically change within a packet transmission, thereby
requiring the adaptive equalization techniques so far described.
At this point, it would be important to address two additional
issues. The first is to account for the gains of reduced-rank
techniques over full-rank methods when the order of the estimators
changes. Another important aspect to be investigated is the
applicability of the proposed reduced-rank techniques to broadband
communications such as MIMO-OFDM systems \cite{li,stuber}. Even
though in MIMO-OFDM systems the frequency selective channels are
transformed into frequency flat channels, there is still the need
to perform spatial equalization. We consider an experiment with a
MIMO-OFDM system in which the data streams per sub-carrier are
separated by MIMO linear equalizers equipped with full-rank and
reduced-rank algorithms and the channels change at each OFDM
block. The system has $N=64$ sub-carriers and employs a cyclic
prefix that corresponds to $C=8$ symbols. The channel profile is
identical to the model employed for the previous experiments and
the fading is independent for each stream. The $N_R \times 1$
received data vector for the $n$th subcarrier is given by
\begin{equation}
{\boldsymbol r}_n[i] = {\boldsymbol H}_n[i] {\boldsymbol x}_n[i] +
{\boldsymbol n}_n[i], ~~~~ n=1,2, \ldots, N,
\end{equation}
where the $N_R \times N_T$ channel matrix ${\boldsymbol H}_n[i]$
contains the channel frequency response gains at the $n$th tone,
the $N_T \times 1$ data vector ${\boldsymbol x}_n[i]$ corresponds
to the symbols transmitted by the $N_T$ antennas over the $n$th
subcarrier and the $N_R \times 1$ vector ${\boldsymbol n}_n[i]$
represents the noise vector at the $n$th tone.

We employ the proposed MIMO linear equalization scheme for spatial
equalization on a per subcarrier-basis \cite{li} for the OFDM
symbols with the proposed and analyzed reduced-rank estimation
algorithms. The BER is plotted against the number of antennas in a
MIMO-OFDM system that has $N_R=N_T$. The results in Fig. 9 show
that the advantages of reduced-rank algorithms are more pronounced
for larger systems, in which the training requirements are more
demanding in term of training data for full-rank RLS algorithms.

\begin{figure}[!htb]
\begin{center}
\def\epsfsize#1#2{1\columnwidth}
\epsfbox{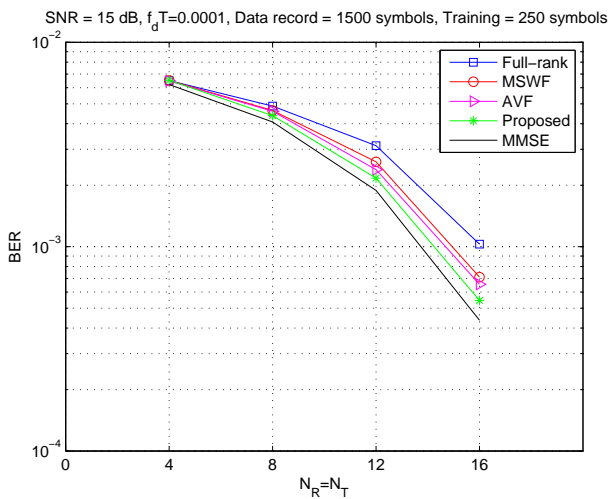} \caption{BER performance versus the number of
antennas .}
\end{center}
\end{figure}

The advantages of the reduced-rank estimators are due to the reduced
amount of training and the relatively short data record (packet
size). Therefore, for packets with relatively small size, the faster
training of reduced-rank LS estimators will lead to superior BER to
conventional full-rank LS estimators. As the length of the packets
is increased, the advantages of reduced-rank estimators become less
pronounced for training purposes and so become the BER advantages
over full-rank estimators. In comparison with the MSWF and AVF
reduced-rank schemes, the proposed scheme exploits the joint and
iterative exchange of information between the transformation matrix
and the reduced-rank estimators, which leads to better performance.
{  The gains of the reduced-rank techniques over full-rank methods
for MIMO-OFDM systems are less pronounced than those observed for
narrowband MIMO systems with multipath channels. This is because the
number of coefficients for estimation is significantly reduced. If
we increase the number of antennas in MIMO-OFDM systems to a large
value, then the gains of reduced-rank techniques become larger.}

\section{Concluding Remarks}

This paper has presented a study of reduced-rank equalization
algorithms for MIMO systems. We have proposed an adaptive
reduced-rank MIMO equalization scheme and algorithms based on joint
iterative optimization of adaptive estimators. We have developed LS
expressions and efficient RLS algorithms for the design of the
proposed reduced-rank MIMO equalizers. A model-order selection
algorithm for automatically adjusting the model order of the
proposed algorithm has also been developed. An analysis of the
convergence of the proposed algorithm has been carried out and
proofs of global convergence of the algorithms have been
established. Simulations for MIMO equalization applications have
shown that the proposed schemes outperforms the state-of-the-art
reduced-rank and the conventional estimation algorithms at a
comparable computational complexity. {  Future work and extensions
of the proposed scheme may consider strategies with iterative
processing \cite{delamaretc,sun,wang} with the aid of convolutional,
Turbo and LDPC codes, detection structures which attain a higher
diversity order \cite{delamaretc,choi} and their theoretical
analysis. }

\end{document}